\documentclass[10pt]{article}
\pdfoutput=1
\usepackage{graphicx,amssymb,amsfonts,amsmath,amssymb,amscd,amstext, mathrsfs,color}
\makeatletter \@addtoreset{equation}{section} \makeatother

\newcommand{\be}{\begin{eqnarray}}
\newcommand{\ee}{\end{eqnarray}}
\newcommand{\ba}{\begin{array}}
\newcommand{\ea}{\end{array}}
\newcommand{\bal}{\begin{align*}}
\newcommand{\eal}{\end{align*}}

\newcommand{\nn}{\nonumber}

\renewcommand{\(}{\Big(}
\renewcommand{\)}{\Big)}
\renewcommand{\[}{\Big[}
\renewcommand{\]}{\Big]}
\def \<{\langle}
\def \>{\rangle}
\definecolor{ggg}{rgb}{0,.6,0}
\begin{document}
 \vspace{0.5cm}
\begin{center} {\Large \bf  OTOC and Quamtum Chaos of   Interacting   Scalar Fields   }
\\
                                            
\vspace{1cm}

Wung-Hong Huang*\\
\vspace{0.5cm}
Department of Physics, National Cheng Kung University,\\
No.1, University Road, Tainan 701, Taiwan\\
                      
\end{center}
\vspace{1cm}
\begin{center} {\large \bf  Abstract} \end{center} 
Discretizing the  $\lambda \phi^4$ scalar field theory on a lattice yields  a system of coupled anharmonic oscillators  with quadratic and quartic potentials. We  begin by analyzing   the two coupled  oscillators in the second quantization method to derive several  analytic relations to the second-order perturbation, which are then employed to numerically calculate the  thermal out-of-time-order correlator  (OTOC), $C_T(t)$. We find  that the function $C_T(t)$ exhibits   exponential growth over a long time window  in the early stages, with Lyapunov exponent $\lambda\sim T^{1/4}$,  which diagnoses quantum chaos.  We furthermore investigate the quantum chaos properties in a closed chain of  N coupled  anharmonic oscillators, which relates to the 1+1 dimensional interacting quantum scalar field theory.  The results reveal an interesting property that the  signatures of quantum chaos appear at low perturbative orders in the OTOC.
\\
\\
\\
\\
\\
\\
\\
\begin{flushleft}
* Retired Professor of NCKU, Tainan, Taiwan. \\
* E-mail: whhwung@mail.ncku.edu.tw
\end{flushleft}
\newpage
\tableofcontents

\section{Introduction}
The exponential growth of out-of-time-order correlator (OTOC) was first discussed by Larkin and Ovchinnikov  \cite{Larkin} to study superconductor many years ago.  Kitaev \cite{Kitaev15a,Kitaev15b,Sachdev} recently revived the concept for studying the SYK model, sparking broad interest across physics fields, including condensed matter and high-energy physics. The function of out-of-time-order correlator (OTOC)  is defined by
\be
 C_T(t)=-\<[W(t),V(0)]^2\>_T   \sim  e^{2\lambda t}
\ee
To see the physical property of $\lambda$  in the exponential function we  consider the case with  W(t) = x(t)  and V = p. In the    classical-quantum correspondence  the commutation relation is replaced by Poisson bracket. : $[A, B]/i\hbar\to \{A,B\}$   and      $C_T(t)  =\hbar^2({\partial x(t)\over \partial x(0)})^2 $.  The Lyapunov exponent $\lambda$ is befined by $|{\partial x(t)\over \partial x(0)}|\sim  e^{\lambda t}$, which  measures  the sensitivity to initial conditions and the quantum OTOC  grows as $\sim  e^{2\lambda t}$. Therefor, the quantum Lyapunov exponent $\lambda$ can be directly extracted from the OTOC.

After the discovery that the Lyapunov exponent in quantum chaos saturates the bound proposed in works by Maldacena et al. \cite{Maldacena15, Maldacena16, Kitaev15c}, many researchers began investigating related problems using conformal field theory (CFT) and AdS/CFT duality tools  \cite{Shenker13a, Shenker14, Roberts14, Shenker13b, Susskind, Liam, Verlinde, Kristan}.

The quantum mechanical method of calculating OTOC  with general Hamiltonian  was set up by Hashimoto recently in  \cite{Hashimoto17,Hashimoto20a,Hashimoto20b}.  For  simple  harmonic oscillator (SHO) the exact OTOC can be easily found and is a purely oscillation function.   Using this method, many complex examples were examined, including the two-dimensional stadium billiard \cite{Hashimoto17,Rozenbaum2019}, the Dicke model \cite{Chavez-Carlos2018}, and bipartite systems \cite{Prakash2020,Prakash2019}.  These models exhibit classical chaos, characterized by exponential growth in OTOCs at early times followed by saturation at late times. The method has also been applied to various systems, including those in many-body physics (e.g., \cite{Jahnke,Das, Romatschke,Shen,Swingle-a, Cotler, Rozenbaum,Bhattacharyya, Morita}).
\\

In the Hashimoto approach the properties of  OTOC  are found by using numerical method in which the first step is to find the wave function of the states therein.  In a  previous unpublished  note  \cite{Huang2306} we use the analytic method  of perturbation in second  quantization approach to study OTOC of  a harmonic  oscillator with extra anharmonic (quartic) interation\footnote{The reference \cite{Romatschke} studied the OTOC of oscillators with pure quartic interaction in wavefunction approach. The system has exact solution of wave function and spectrum.}. Our method offers the advantage of directly determining the properties of any quantum level "n," while the wavefunction approach necessitates a step-by-step numerical evaluation for each quantum level "n" to extract its properties.

According to our method, to the first order pertubation  \cite{Huang2306}, however, we does not see the exponential growth in the initial time nor the saturation to a constant OTOC in the final times, i.e. $ C_T(\infty)\to 2\<x^2\>_T\<p^2\>_T$,   which are known to associate with quantum chaotic behavior in systems that exhibit chaos \cite{Maldacena16}.  In the next  note  \cite{Huang2311} we extended the method to the second-order perturbation and found that OTOC  saturates to a constant value at later times. However, at early times, the OTOC will increase rapidly following a quadratic power law, rather than exhibiting the exponential growth that is essential for the emergence of chaotic dynamics. In the  third of a series of our study  \cite{Huang2407} we showed that in systems with sufficiently strong quartic interactions, an exponential growth curve may emerge at third-order perturbation, although this is not yet certain.

This paper is the fourth in a series of our studies on the perturbative OTOC using the second quantization approach.  This time, we turn to study the second-order OTOC of interacting quantum scalar field theory. We will see that the  OTOC  saturate to a constant value at later times and show the  exponential growth in the early stage, which diagnose the quantum chaos. 
\\

 In section 2, we  regularize the  interacting quantum $\phi^4$ scalar field theory by placing it on a square lattice and see that the theory becomes a quantum mechanical system of coupled anharmonic oscillators, in which  the  anharmonic oscillator describes a simple harmonic oscillator with extra quartic potential. For self-consistency, we also provide a brief review of Hashimoto's method for computing quantum mechanical OTOCs. 

  In section 3  we  use the   second quantization method to calculate the  OTOC in the systems  of  coupled anharmonic oscillators.  We obtain the analytic relations  of spectrum, Fock space states and matrix elements of coordinate. 

In Section 4,  by using these relations, we numerically evaluate OTOC $C_T(t)$  . We plot several diagrams   to see that the function $C_T(t)$ exhibits the  exponential growth  fitting over a long time window  in the early stages with Lyapunov $\lambda\sim T^{1/4}$,  which diagnose the quantum chaos. 

In Section 5 we use the found property of   coupled anharmonic oscillators  to analyze the closed  chain of 3 and 4 coupled anharmonic oscillators and find the quantum chaos therein.  We then argue that the quantum chaos property also shows in the 1+1 dimention interacting  scalar field theory.    The final section provides a brief discussion.
\section{Interacting scalar fields and coupled  aharmonic oscillators}
\subsection{Hamiltonian of lattice  $\lambda \phi^4$ theory}
The Hamiltonian   $d$-dimensional massive scalar  with  $\hat\lambda \phi^4$ interaction  is
\be
{\cal H}=\frac{1}{2}\int d^{d-1}x\left[\pi(x)^2+\vec\nabla\phi(x)^2+m^2\phi(x)^2+{\hat\lambda\over 12}\phi(x)^4\right] \ .
\ee
To calculate the OTOC of the above integrating scalar field theory we place the theory on a square  lattice with lattice spacing  $\delta$, then, with $\cdot{\cal H}= \delta^{d-1}\cdot{ H}$
\be
H=  \sum_{\vec n}\left\{{1\over2}  {p(\vec n)^2 }+\sum_i{1\over 2 \delta^2} \(\phi(\vec n)-\phi(\vec n-\hat{a}_i)\)^2+{1\over2}m^2\phi(\vec n)^2+{\hat\lambda\over 24}\phi(\vec n)^4 \right\} \ ,
\ee
where $\hat{a}_i$ are d-dimensional unit vectors pointing toward the spatial directions of the lattice and $\vec n$ is the position of lattice point.  We can  then define 
\be
&& X(\vec n)={\phi(\vec n)\over \delta^{d/2}}, ~~~ P(\vec n)={p(\vec n) \over \delta^{d/2}}, ~~ M={1\over \delta}, ~~\tilde \omega=  m\ \delta^{ 2}, ~~~ \Omega={ \delta^2}, ~~~ \lambda={\hat \lambda \over 24}\ \delta^{3d-2} 
\ee 
to obtain the lattice Hamiltonian of $H= \sum_{\vec n}\ H_{\vec n}$  with
\be
 H_{\vec n}= \left\{\frac{P(\vec n)^2}{2M}+\frac12 M \left[\tilde\omega^2 X(\vec n)^2+\Omega^2\sum_i\Big( X(\vec n)-X(\vec n-\hat{a}_i)\Big)^2+2\lambda\,X(\vec n)^4\right]\right\} \label{HH}
\ee 
When $\vec n$ is an one dimensional vector, the Hamiltonian describes an infinite family of coupled d-1 dimensional oscillators.

  In  the simplest case of 2 coupled oscillators, we can  define the new coordinate
\be
X_1\to {1\over \sqrt 2}(x_1+x_2),~X_2\to {1\over \sqrt 2}(x_1-x_2)  \\
P_1\to {1\over \sqrt 2}(p_1+p_2),~P_2\to {1\over \sqrt 2}(p_1-p_2) 
\ee
and  Hamiltonian becomes
\begin{align}
H &=  {1\over 2}\( p_1^2+ \omega ^2 x_1^2+ p_2^2+ (\omega ^2+2\Omega ^2) x_2^2\)+{\lambda\over 2}\,\(x_1^4+ x_2^4+6 x_1^2 x_2^2\)     \nn\\
&=   {1\over 2}\( p_1^2+ \omega_1 ^2 x_1^2+ p_2^2+\omega_2 ^2 x_2^2\)+{\lambda\over 2}\,\(x_1^4+ x_2^4+6 x_1^2 x_2^2\)   =K+V,~~~\omega_1=\omega,~\omega_2 ^2 =\omega ^2+2 \Omega ^2    \label{V1}
\end{align}
In this paper we will use above Hamiltonian to analyze the quantum chaos therein. We will see that  the temperature dependence of the Lyapunov  is $\lambda_T\sim T^{1\over4}$. With the property we will show that the linear closed chain, which is related to 1+1  dimensional interacting quantum scalar field theory shows quantum chaos too.

The interaction form in eq.(\ref{V1})  tells us that the Hamiltonian  describes two coupled anharmonic oscillators, in which each one is just the harmonic oscillator with extra quartic potential (${\lambda\over 2}x_1^4$ and ${\lambda\over 2} x_2^4$) and coupled interaction is  ${3\lambda } x_1^2x_2^2$.  See the figure 1.
\\

\scalebox{0.3  }{\hspace{10cm}\includegraphics{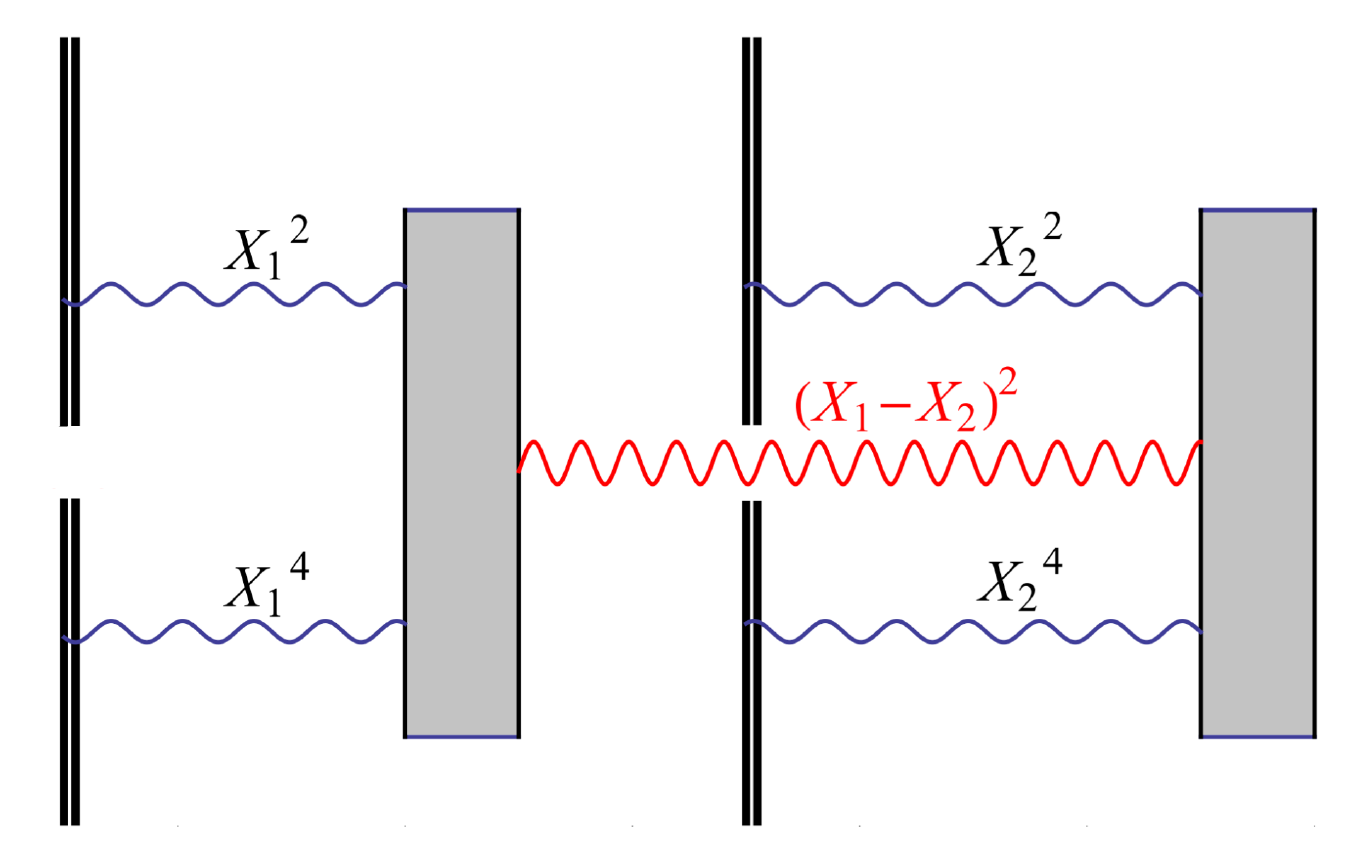}}
\\

{Figure 1:    Two coupled anharmonic oscillators with quaratic potential ($ X_1^2, X_2^2$), quartic potential ($ X_1^4,~ X_2^4$) and coupled interaction $  (X_1-X_2)^2$. }
\\

 Above interaction term  was used by us in \cite{Huang21} to study the complexity of interacting sclar field theory. 
Note that Removing $x_1^4$ and $x_2^4$ terms in above Hamiltonian will describe the system of  {\it two  coupled simple harmonic oscillators},  which is known to be obtained by a reduction from SU(2) Yang-Mills-Higgs theory.  The associated OTOC and the quantum chaos properties  was studied by  Hashimoto  \cite{Hashimoto20a} using the wavefunction approach\footnote{Note that the Lyapunov of matrix $\Phi^4$ theory has been studied by  Stanford \cite{Stanford} in many years ago and more recently by   Kolganov \cite{Kolganov}.}.   

This paper is to investiage the OTOC of the two couple anharmonic oscillators. As the systems describe the quantum mechanical modes we can use  the Hashimoto's method to  calculate  the asociated OTOC.  For the self-consistency, we will briefly summarize the method in the next subsection.  
\subsection{Hashimoto method to OTOC : second quantization method}
 For a  time-independent Hamiltonian: $H = H(x_1,....x_n,p_1,....p_n)$ the function of  OTOC is  define by 
\be
 {C_T(t)={1\over Z}\sum_ne^{-\beta E_n}\,c_n(t)} ,~~~c_n(t)\equiv -\<n|  [x(t),p(0)]^2         |n\>    \label{TC}
\ee
where   $|n\>$ is the energy eigenstate.   We  first insert  the  complete set $\sum_m|m\>\<m|=1$ to find a relation
\be
  { c_n(t)}  & =& -\sum_m\<n|  [x(t),p(0)]|m\>\<m|  [x(t),p(0)] |n\>\    { =   \sum_m(ib_{nm})(ib_{nm})^*} ~\label{cn}\\
b_{nm}&=& -i\<n|  [x(t),p(0)]|m\>,~~b_{nm}^*=b_{mn} 
\ee
After using relation $ x(t) = e^{ iHt/\hbar  }\,x\,e^{- i  Ht/\hbar} $ and inserting the completeness relation again  we obtain
\be
 b_{nm}&\equiv& -i\<n|  x(t), p(0)|m\> +i\<n| p(0) x(t),|m\>\nn\\
&=& -i\sum_k\(e^{i  E_{nk}t/\hbar}x_{nk}p_{km}-e^{i E_{km}t/\hbar}p_{nk}x_{km}\) \\
E_{nm}&=& E_n-E_m,~~~ x_{nm}=\<n| x|m\> ,~~p_{nm}=\<n|p |m\>~~~~\label{Enm}
\ee
We are interesting in  the  quantum mechanical Hamiltonian\footnote{{ Notice that Hashimoto \cite{Hashimoto17} used $H=\sum_i{p_i^2}+U(x_1,....x_N)$ which is that in our notation for M=1/2.  Therefore the formula $b_{mn}$ in eq.(\ref{bnm})  becomes Hashimoto's   formula if M=1/2 and $\hbar=1$.}}
\be
&&H=\sum_i{p_i^2\over 2M}+U(x_1,....x_N)~~\to~~[H,x_i] = - i\hbar { p_i\over M} 
\ee
where  $M$ is the particle mass. Using the relations
\be
p_{km}&=&\<k|p |m\>={iM\over \hbar }\<k|[H,x] |m\>={iM\over \hbar }\<k|(E_k\,x)-(x\,E_m) |m\>={iM\over \hbar }(E_{km})x_{km} 
\ee
we have a simple formula
\be
 { b_{nm} = { M\over \hbar }\sum_k\,x_{nk}x_{km}\(e^{i  E_{nk}t/\hbar} E_{km} -e^{i  E_{km}t/\hbar}E_{nk}\)} ~~~~~~~\label{bnm}
\ee
Now  we can compute OTOC through (\ref{bnm}) once we know $x_{nm}$ and $E_{nm}$ defined in (\ref{Enm}).  

While the original method of  \cite{ Hashimoto17} used the wavefunction approach we will calculate the OTOCs   by the perturbation in the second quantization approach. In this approach  the  kinetic term has a diagonal form 
\be
K &=&\omega_1 a_1^\dag  a_1+\omega_2 a_2^\dag  a_2+{1\over 2} (\omega_1+\omega_2) \label{K}
\ee
 and  interaction term is
\be
V&=& {\lambda \over 8}\left[\(\sqrt{1\over \omega_{1}}( a_{1}^\dag+ a_{1})\)^4+\(\sqrt{1\over \omega_{2}}( a_{2}^\dag+ a_{2})\)^4+6\(\sqrt{1\over \omega_{1}}( a_{1}^\dag+ a_{1})\)^2\(\sqrt{1\over \omega_{2}}( a_{2}^\dag+ a_{2})\)^2\right]   \label{V}
\ee
which will be regarded as a perturbation in later calculation.
 
Consider first the case of V=0. The system describes two uncoupled simple harmonic oscillators, and  each Hamiltonian $H$, state $|n\>$, spectrum $E_n$ and $E_{nm}$ are
\be
  H&=&  \hbar\omega \( a^\dag a+ \frac{1}{2}\),~~H|n\>=E_n|n\>,~~E_n=\hbar\omega\left(n+\frac{1}{2}\right)
\ee
Basic  relations
\be
x|n\>=\sqrt{\hbar\over2M\omega}(a^\dag+a ) |n\>=\sqrt{\hbar\over2M\omega}\ \sqrt{n}|n-1\>+  \sqrt{\hbar\over2M\omega}\ \sqrt{n+1} |n+1\>
\ee
quickly leads to
\be
x_{nm}&\equiv&\<n|x|m\>=\sqrt{\hbar\over2M\omega}\( \sqrt{m}\ \delta_{n,m-1}+   \sqrt{m+1}\ \delta_{n, m+1}\)   \label{SQxnm}
\ee
Substituting above expressions into  (\ref{Enm}) and  (\ref{bnm}) we obtain 
\be 
 b_{nm}(t)&=&{ M\over \hbar }\sum_k\,x_{nk}x_{km}\(e^{i  E_{nk}t/\hbar} E_{km} -e^{i E_{km}t/\hbar}E_{nk}\)
=   \hbar \cos (\omega t)\ \delta_{nm}
\ee
Then
\be
c_n(t)&=&   \hbar^2  \cos^2 (  \omega t),~~~C_T(t)=  \hbar^2 \cos^2(  \omega t) ~~~~~~~~~~~\label{HR}
\ee
Both of $c_T(t)$ and $C_T(t)$ are periodic functions and   do not depend on energy level $n$ nor temperature $T$.  This distinctive property of the harmonic oscillator was first emphasized in the original paper \cite{Hashimoto17}. In the next section, we employ second quantization to calculate the second-order OTOC for coupled anharmonic oscillators, showing that quantum chaos emerges within the perturbative approximation.
\section{Second-order OTOC of  coupled anharmonic oscillators  : analytic relations}
As described in eq.(\ref{K}) and eq.(\ref{V}) the Hamiltonian of the coupled anharmonic oscillators we  considered is, after introducing the parameter $\eta$ 
\be
H&=&H^{(0)}+V,~~~~~~   \nn\\
&=&\omega_1 a^\dag_1a _1+\omega_2 a^\dag_2a _2+ {\lambda \over 8}\left[\eta\({1\over \omega_{1}^2}( a_{1}^\dag+ a_{1})^4+{1\over \omega_{2}^2}( a_{2}^\dag+ a_{2})^4\)+6\(\ {1\over \omega_{1}}( a_{1}^\dag+ a_{1}) ^2{1\over \omega_{2}}( a_{2}^\dag+ a_{2}) ^2\)\right]\nn\\
\ee
where $H^{(0)}$ describes the  kinetic energy of the simle harmnic oscillator.  In the  case of  $\eta=1$ the above equation describes the coupled anharmonic oscillators. When $\eta=0$ it describes the coupled simple  harmonic oscillators which was studied in reference \cite{Hashimoto20a} by the wavefunction approach.  Therefore, our analysis could reproduce their results while in the second quntization method. That is, using the above Hamiltonian, we can study the perturbative OTOC for coupled anharmonic and coupled simple harmonic oscillators in a unified form.

In this section, we calculate the second-order perturbative OTOC for the system described above. Note that the first-, second-, and third-order calculations for the {\it uncoupled anharmonic oscillator} were completed in our unpublished notes~\cite{Huang2306,Huang2311,Huang2407}, which, however, do not exhibit quantum chaos. 

The coupling term $x_1^2x_2^2$ in  here makes the system more complex which, however, could induce quantum chaotic properties in the second-order perturbation.
\subsection{Perturbative energy and state :  text formulas}
The coupled  anharmonic  oscillators has a well-known unperturbed solution  (set $M=\hbar=1$)
\be
&&H^{(0)} |n_1^{(0)},n_2^{(0)} \>=E^{(0)}_n|n_1^{(0)},n_2^{(0)} \>=\( \omega_1 n_1^{(0)}+ \omega_2n_2^{(0)} +{1 }\) |n_1^{(0)},n_2^{(0)} \>  \label{3.2}
\ee
Hereafter  we will use notation
\be
&&|   n   \>=| \vec n  \>=|n_1 ,n_2  \>;~~|   n^{(0)} \>=| \vec n^{(0)} \>=|n_1^{(0)},n_2^{(0)} \>;~~|   n^{(1)} \>=| \vec n^{(1)} \>=|n_1^{(1)},n_2^{(1)} \>
\ee
as the   short symbol without the confusion.

The second-order perturbative  energy and the state formulas in quantum mechanics  are 
\be
E_n &= & E_n^{(0)} +  E_n^{(1)} +   E_n^{(2)}+{\cal O}(\lambda^3) \\
 |n\>&=& |n^{(0)} \>+  |n^{(1)} \>+\lambda^2 |n^{(2)} \> +{\cal O}(\lambda^3)            \label{n}
 \ee
where
\begin{align}
E^{(1)}_n&=\lambda   \<n^{(0)} | V   |n^{(0)}  \>  \label{3.6}\\
E^{(2)}_n&=\lambda^2 \sum_{k\ne n}{| \<k^{(0)} |V|n^{(0)} \> |^2   \over E^{(0)}_n -E^{(0)}_k} \\
  {|n^{(1)}\>}&= \lambda  \sum_{k\ne n}|k^{(0)} \>{\<k^{(0)} |  V    |n^{(0)} \> \over E^{(0)}_n -E^{(0)}_k } \label{3.7}\\
{|n^{(2)}\>}&=  -{\lambda^2 \over2}  |n^{(0)}  \>\sum_{k\ne n}{| \<k^{(0)} |V|n^{(0)} \> |^2   \over (E^{(0)}_n -E^{(0)}_k)^2}   \nn   \\
& - \lambda^2 \sum_{k\ne n}|k^{(0)} \>{\<k^{(0)} |  V    |n^{(0)} \>  \<n^{(0)} |  V    |n^{(0)} \> \over (E^{(0)}_n -E^{(0)}_k )^2}+ \lambda^2  \sum_{k\ne n}\sum_{\ell\ne n}|k^{(0)} \>{\<k^{(0)} |  V    |\ell^{(0)} \> \over E^{(0)}_n -E^{(0)}_k }{\<\ell^{(0)} |  V    |n^{(0)} \> \over E^{(0)}_n -E^{(0)}_\ell } 
\end{align}
In the remainder of this section, we apply the above formulas to calculate:

1.  Perturbative Energy   $  E$. 

2. Perturbative state $ {|n\>}$. 

3. Perturbative matrix elements $ {x_{mn}}$. 
\\
Using these analytic results, we calculate the OTOC and analyze its properties in the next section.
\subsection{Perturbative $V_{kn}$ and perturbative energy $  E_n$  : model calculations}
 A crucial quantity we need, after calculation, is  (we let parameter $\eta=1$ hereafter)
\be
&&\boxed{V_{kn}=\<k^{(0)})|V|n^{(0)}\>=\<k_1^{(0)},k_2^{(0)})|{\lambda\over2}(   x_1^4+  x_2^4+6x_1^2x_2^2)|n_1^{(0)},n_2^{(0)}\>}   \\
&=& {\lambda\over 8}\(     \sqrt{ n_2 +1} \sqrt{ n_2 +2} \sqrt{ n_2 +3} \sqrt{ n_2 +4}\   \delta_{ k_1,  n_1}  \,\delta_{ k_2, n_2 +4}              + 6 \sqrt{ n_1+1} \sqrt{ n_1+2} \sqrt{ n_2 +1} \sqrt{ n_2 +2}\,  \delta_{ k_1, n_1+2}  \,\delta_{ k_2, n_2 +2}                                      \nn  \\
&& + 2 \sqrt{ n _2    +1} \sqrt{ n _2    +2} (6  n _1    +   (2  n _2    +3)+3)\, \delta_{ k_1, n_1}\, \delta_{ k_2, n_2 +2} +   6 \sqrt{ n_1-1} \sqrt{ n_1} \sqrt{ n_2 +1} \sqrt{ n_2 +2} \, \delta_{ k_1, n_1-2} \, \delta_{ k_2, n_2+2}                                             \nn  \\
&& +    \sqrt{ n_1+1} \sqrt{ n_1+2} \sqrt{ n_1+3} \sqrt{ n_1+4}\   \delta_{ k_1, n_1+4}  \,\delta_{ k_2, n_2}               + 2 \sqrt{ n _1    -1} \sqrt{ n _1    } (\eta  (2  n _1    -1)+6  n _2    +3)    \, \delta_{ k_1, n_1+2} \, \delta_{ k_2, n_2}              \nn           \\
&&+  6 \left(\eta (1+ n _1^2)+ n _1     ( 4  n _2    +2)+ n _2     (  (1+ n _2)    +2)+1\right) \, \delta_{ k_1, n_1}\,  \delta_{ k_2, n_2 } \nn\\
&& +2 \sqrt{ n _1    -1} \sqrt{ n _1    } (\eta  (2  n _1    -1)+6  n _2    +3) \, \delta_{ k_1, n_1-2} \, \delta_{ k_2, n_2}\nn\\
&& +    \sqrt{ n_1-3} \sqrt{ n_1-2} \sqrt{ n_1-1} \sqrt{ n_1}\    \delta_{ k_1, n_1-4}\, \delta_{ k_2, n_2 }  +     6 \sqrt{ n_1+1} \sqrt{ n_1+2} \sqrt{ n_2 -1} \sqrt{ n_2 }\,  \delta_{ k_1, n_1+2}\,  \delta_{ k_2, n_2 -2}                                         \nn  \\
&& + 2 \sqrt{ n _2    -1} \sqrt{ n _2    } (6  n _1    +  (2  n _2    -1)+3) \delta_{ k_1, n_1} \, \delta_{ k_2, n_2 -2} +     6 \sqrt{ n_1-1} \sqrt{ n_1} \sqrt{ n_2 -1} \sqrt{ n_2 } \, \delta_{ k_1 , n_1-2}\,  \delta_{ k_2, n_2 -2} \nn  \\
&&+   \sqrt{ n_2 -3} \sqrt{ n_2 -2} \sqrt{ n_2 -1} \sqrt{ n_2 }\   \delta_{ k_1, n_1}\,\delta_{ k_2, n_2-4}\)
\ee
Using above result the second-order perturbative   energy $ {E_n}$  becomes
\be
E_n &= & E_n^{(0)} +  E_n^{(1)}  + E_n^{(2)}  +{\cal O}(\lambda ^2) \\
 {E^{(0)}_n} &=&\omega_1 \(n_1+{1\over2}\)+ \omega_2 \(n_2+{1\over2}\)  \\
{E^{(1)}_n} &=& {3 \lambda  \over4}\(1+2(n_1+n_2)+4n_1 n_2+ (1+n_1+n_1^2+n_2+n_2^2) \) \\
{E^{(2)}_n} &=&\frac{-9 \lambda^2}{16 \left(\omega _1+\omega _2\right)}       (1+  n _1   +   n _2) (2+   n _1   +   n _2 +2    n _1       n _2     )   
\ee
\subsection{Perturbative  state $ {|n\>}$   :  model calculations}
To proceed we   calculate the perturbative  states of $  {|n\>} $.  The results, with a notation $|n^{(j)}\>=|n \>^{(j)}$, are
\be
&\bullet&{|n\>}= |n^{(0)} \>+  |n^{(1)} \> +  |n^{(2)} \> +{\cal O}( \lambda^3) \\
&\bullet&|n\>^{(1)}\nn\\
&=&{3 \lambda\over 8}\( \frac{ \sqrt{ n_1   -1} \sqrt{ n_1   }\sqrt{ n_2  -1} \sqrt{ n_2  }  }{\omega _1+\omega _2}\, |n_1-2,n_2-2\>^{(0)}   \nn\\
&&+ \frac{ \sqrt{ n_1   -1} \sqrt{ n_1   }  \sqrt{ n_2  +1} \sqrt{ n_2  +2} }{\omega _1-\omega _2} |n_1-2,n_2+2\>^{(0)} \nn\\
&&- \frac{\sqrt{ n_1   +1} \sqrt{ n_1   +2}\sqrt{ n_2  -1} \sqrt{ n_2  } }{\omega _1-\omega _2}\, |n_1+2,n_2-2\>^{(0)} \nn\\
&&-\frac{\sqrt{ n_1   +1} \sqrt{ n_1   +2}\sqrt{ n_2  +1} \sqrt{ n_2  +2} }{\omega _1+\omega _2}  |n_1+2,n_2+2\>^{(0)} \)
 \label{1}\\
&\equiv&f(n_1,n_2)_{-2,-2}\, |n_1-2,n_2-2\>^{(0)}+f(n_1,n_2)_{-2, 2}\, |n_1-2,n_2+2\>^{(0)}\nn\\
&&+f(n_1,n_2)_{ 2,-2}\, |n_1+2,n_2-2\>^{(0)}+f(n_1,n_2)_{2,2}\, |n_1+2,n_2+2\>^{(0)}                                    \label{nn} \\
&\bullet&|n\>^{(2)} \nn\\
&=&\frac{3\lambda^2}{128}\(\frac{3 \sqrt{ n _1    -3} \sqrt{ n _1    -2} \sqrt{ n _1    -1} \sqrt{ n _1    } \sqrt{ n _2    -3} \sqrt{ n _2    -2} \sqrt{ n _2    -1} \sqrt{ n _2    }  }{\left(\omega _1+\omega _2\right){}^2} |n_1-4,n_2-4\>^{(0)}\nn\\
&&+\frac{2 \sqrt{ n _1    -3} \sqrt{ n _1    -2} \sqrt{ n _1    -1} \sqrt{ n _1    } \sqrt{ n _2    -1} \sqrt{ n _2    } (\eta  (2  n _1    -5)+6  n _2    -9)  }{\left(\omega _1+\omega _2\right) \left(2 \omega _1+\omega _2\right)} |n_1-4,n_2-2\>^{(0)}~~~~\nn\\
&&+\cdot\cdot\cdot\ee
We introduce function $f(n_1,n_2)_{k_1,k_2}$  in eq.(\ref{nn}) for later use. Note that the first-order corrected state $|n\>^{(1)}$  has 4 terms while second-order corrected state  $|n\>^{(2)}$ which has 16 terms and the   equation above is not fully written out.
\subsection{Perturbative  matrix elements $ {x_{mn}}$ : model calculations}
Use above relations we could now begin to  calculate the   matrix elements $ {x_{mn}}=\<m|x|n\>$. To second order of $\lambda$ we use the real operator $x_i$
\be
x_i|n\>&=&\sqrt {{1\over 2\omega_i}}(a_i^\dag+a_i)|n\> ,~~~~~~~i=1,2       \label{3.24}
\ee
to calculate
\be
 & & {\<m|x_i|n\> = \left({}^{(0)}\<m|+ ~{}^{(1)}\<m| + ~{}^{(2)}\<m|   \right) \ x_i\   \(|n \>^{(0)}+   |n \>^{(1)} +   |n \>^{(2)}\)}    \label{Nxmn}   \nn\\
&=&{}^{(0)}\<m|x_i |n \>^{(0)}+{}^{(0)}\<m|\ x_i \ |n \>^{(1)} + {}^{(1)}\<m| \ x_i \ |n \>^{(0)}\nn\\
&&+{}^{(0)}\<m|\ x_i \ |n \>^{(2)} + {}^{(2)}\<m| \ x_i \ |n \>^{(0)}+  {}^{(1)}\<m| \ x_i \ |n \>^{(1)}             \nn  \\
 &=&{}^{(0)}\<m|x_i |n \>^{(0)}+  {}^{(0)}\<m| x_i  |n \>^{(1)}+{}^{(0)}\<n| x_i  |m \>^{(1)} + {}^{(0)}\<m| x_i  |n \>^{(2)}+ {}^{(0)}\<n| x_i  |m \>^{(2)}            \nn  \\
&&+ f(m_1,m_2)_{-2,-2}\, {}^{(0)}\<m_1-2,m_2-2| x_i  |n \>^{(1)} + f(m_1,m_2)_{-2,2}\, {}^{(0)}\<m_1-2,m_2+2| x_i  |n \>^{(1)}             \nn  \\
&&+ f(m_1,m_2)_{2,-2}\, {}^{(0)}\<m_1+2,m_2-2| x_i  |n \>^{(1)} + f(m_1,m_2)_{2,2}\, {}^{(0)}\<m_1+2,m_2+2| x_i  |n \>^{(1)}   \label{xmn}~~~
\ee
Component of $\<m|x_i|n\>$ is too long to be written out explicitly in here. 
\section{Second-order   OTOC of coupled anharmonic oscillators : numerical    $C_T(t)$}
To proceed, we substitute the analytic form  of $ {x_{mn}}$ in (\ref{Nxmn}) to calculate the  function $b_{nm}$ in  (\ref{bnm}). Then use the formula in (\ref{cn}) to calculate the associated  microcanonical OTOC formula $c_{n}(t)$, and finally, use the formula in (\ref{TC}) to numerically evaluate  the thermal  OTOC, $C_T(t)$.  

In this section we analyze the properties  of $C_T(t)$ of the couple anharmonic oscillator in detail.  
Note that the numerical plots   in this paper are the result of selecting the following parameters  : $M= \hbar =1,~\omega=1,\lambda=0.1$.  
\subsection{Mode summation in $C_T(t)$}
Using eq.(\ref{TC}) to evaluate  $C_T(t)$ we have to  sum over the mode index n, where  the summation    is performed over the range $0\le n\le n_F$. We plot  figure 2  to show the   properties of $ C_T(t) $ for T=10  with various cutoff mode number  $n_F$ :  $n_F=10,20,30,40$.\footnote{The function $ C_T(t) $ plotted in figures 2 and 3 are rescaled to region  $0< C_T(t)<10$}  
\\
\scalebox{0.3}{\hspace{12cm}\includegraphics{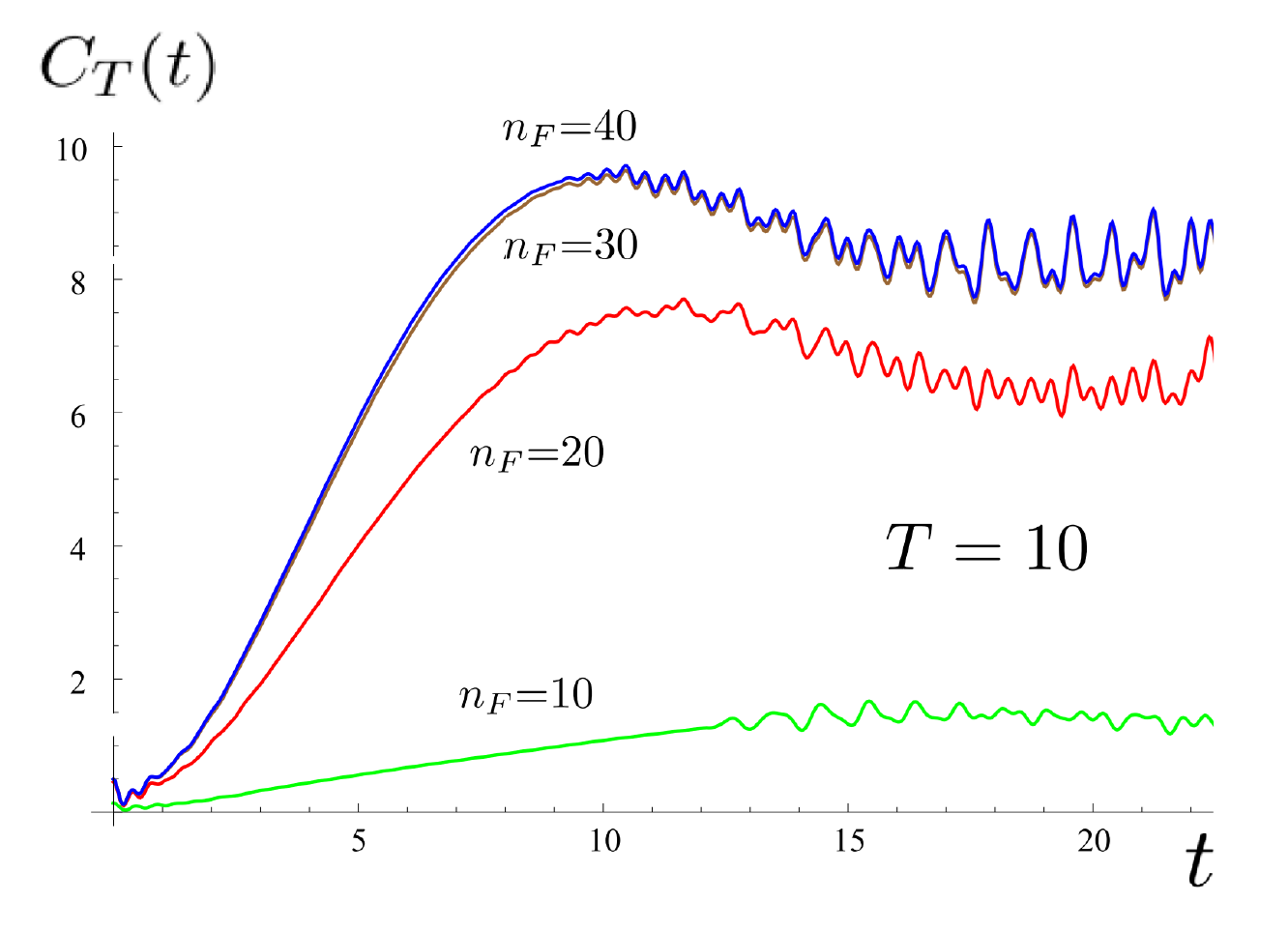}}
\\
{Figure 2:     OTOC  $C_T(t)$  as the function of time for various   cutoff mode number $n_F$. }
\\

We see that $C_T(t)$ for $n_F=30$ is closer to that of $n_F=40$. The property also shows in other temperatures.  Thus the figures 3 and 4 are plotted with $n_F=40$.
\subsection{Exponential growth and Lyapunov exponent}
In Figure 3, we plot   $C_T(t)$,  for a system at temperature T=10 to clearly illustrate that the exponential growth occurs between the dissipation time $t_d\approx 1$ and scrambling time $t_*\approx 5$.  The numerical results yield the Lyapunov exponent $\lambda$ : 
\be 
C_T(t)\sim e^{\lambda\cdot t},~~~\lambda=0.559767  \pm 0.01\%   \label{pl},~~~~~~T=10
\ee
After several numerical calculations we find that the chaotic property   shows in the systems  of $T>6$.
\\
\scalebox{0.25  }{\hspace{15cm}\includegraphics{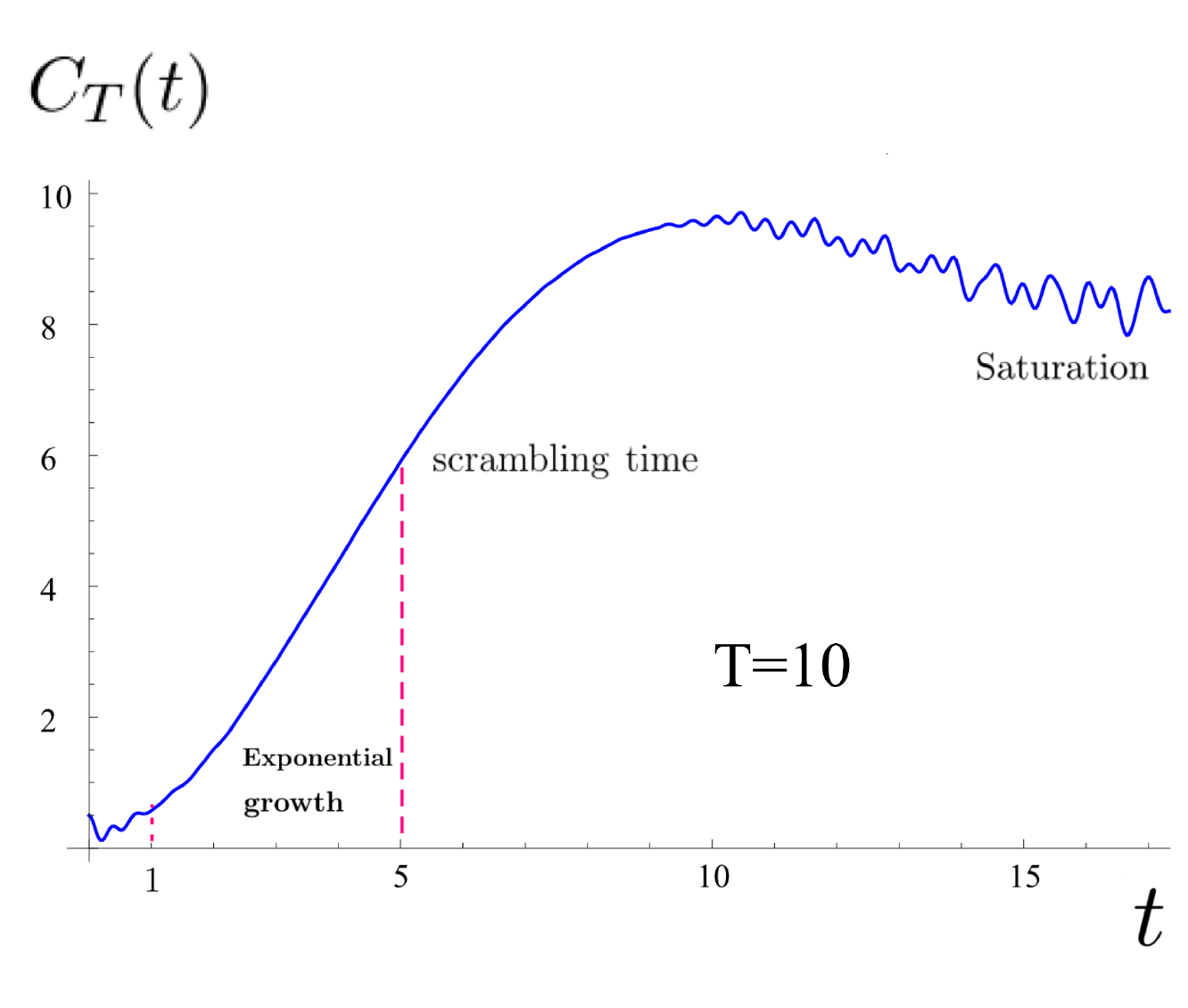}} 
\\
{Figure 3 : The exponential growth of $ C_T(t) $  within  the wide interval $1\le  t \le 5$ for T=10 .}
\\

Note that the value and error of the Lyapunov exponent $\lambda$ in Eq.~(\ref{pl}) depend on the selected time interval. As shown in Figure 3, the initial time is chosen after the dissipation time near $t_d\approx1$, then the curve begins to increase smoothly, and before the scrambling time $t_*\approx5$, at which $C_T(t)$ approaches half of its saturation value  \cite{Jahnke}.

\subsection{Temperature dependence of Lyapunov exponent} \label{sec4.3}
It is interesting to see that the temperature dependence of  Lyapunov in coupled anharmonic oscillator has a simple power law.  
\be
\lambda \sim \kappa\ T^{1/4},~~~~\kappa\sim0.3          \label{pl}
\ee
which is confirmed from the numerical result in Figure 3. The standard errors of Lyapunov exponent obtained from fitting the data, shown in figure 4, are all less than 0.01 \%.  

{Note that the property of  $\lambda\sim T^{1/4}$ in quantum system was first found  in the coupled harmonic oscillators \cite{Hashimoto20a}. The crucial point is that, in  the high energy limit the mass term $x^2+y^2$   can be  ignored,  and energy of oscillators becomes $E=p_x^2+p_y^2+x^2y^2$ which allows the {\it scaling transformation}
\be
(x,y)\to (\alpha x,\alpha y),~E\to \alpha^4E,~t\to \alpha^{-1}t
\ee
The property that Lyapunov exponent has the dimension of inverse time leads to $\lambda\sim E^{1/4}$.

 For the system of coupled anharmonic oscillators considered in this paper the   energy becomes $E=p_x^2+p_y^2+x^2y^2+x^4+y^4$ which has the {\it same scaling transformation} and thus the same relation   eq.(\ref{pl}).}
\\
\scalebox{0.5}{\hspace{4.5cm}\includegraphics{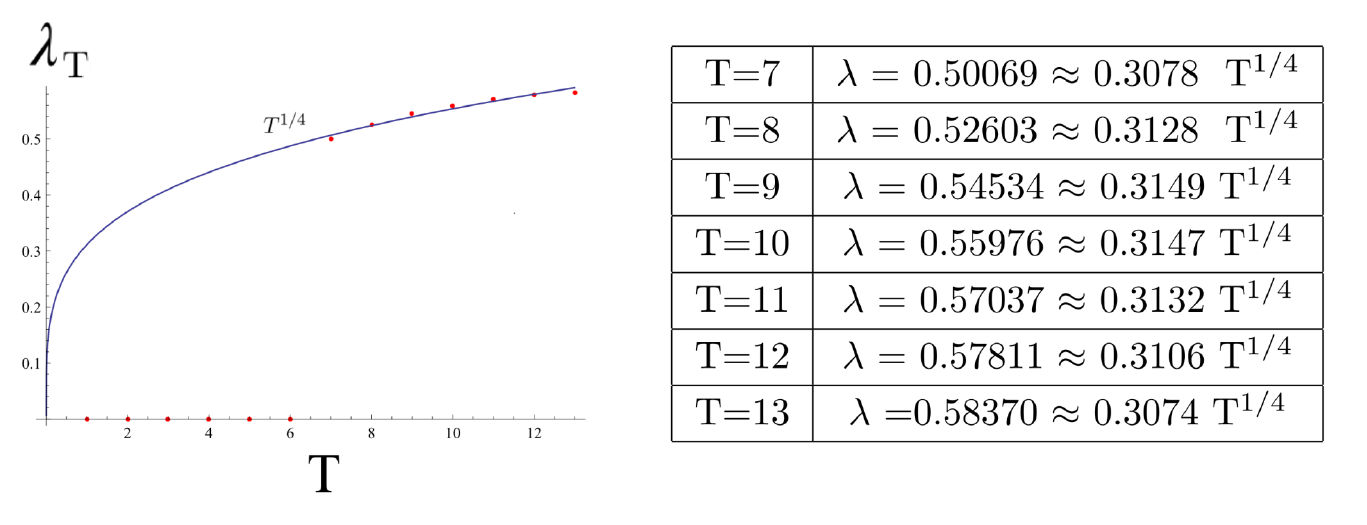}} 
 
\hspace{3cm}Figure 4 : Temperature dependence of  Lyapunov exponent  $\lambda_T$.
\\

Note that the property $\lambda_T\sim T^{1/4}$ satisfies the MSS bound condition  \cite{Maldacena15}.   Above arguments  also implies that, for  for $\lambda \phi^n$ scalar field theory, $\lambda_T\sim T^{1/n}$.
\\

In conclusion, the OTOC behavior of coupled anharmonic oscillators is : After dissipation time near $t_d\approx1$,  it begin to increase by   exponentially growth  up until $t_*\approx 5 $. Then, after scrambling until $t\approx15$, it approaches a constant value with slight oscillations, as shown in Figures 2 and 3. Note that, to first-order perturbation theory, there is no exponential growth and no asymptotic approach to a constant value.

\section{OTOC of interacting quantum scalar fields : closed chain  of  N coupled anharmonic  oscillators}
After studying the OTOC of  the two coupled anharmonic oscillators we now turn to   investigate  the OTOC of an one-dimensional closed chain of  N coupled  anharmonic oscillators which related to the 1+1 dimensional interacting quantum scalar field system.  The analysis can be illustrated by the simplest case of a closed chain of  three coupled   anharmonic oscillators.
\subsection{Closed chain of three  coupled anharmonic oscillators}
From eq.(\ref{HH}) we have the Hamiltonian  (in units where $M=1$) reads
\begin{align}
H&={1\over 2}\Big[\( P_1^2+P_2^2+P_3^2 \)+\tilde \omega^2(X_1^2+ X_2^2+X_3^2 )  + \Omega ^2\(   ( X_1- X_2)^2+ ( X_2- X_3)^2+ ( X_3- X_1)^2 \) \nn\\
& +2\lambda( X_1^4+X2^4+X_3^4 )\Big]    \label{5.1}
\end{align}   
 which describes a closed chain of three  coupled anharmonic oscillators shown in the figure 5.
\\

\scalebox{0.5 }{\hspace{2cm}\includegraphics{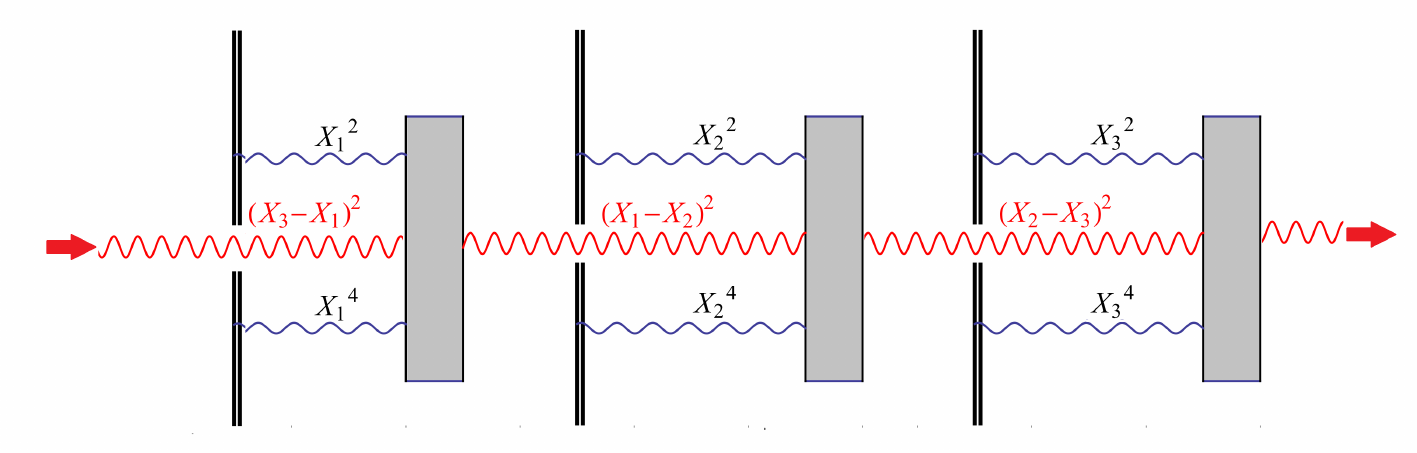}}
\\
{Figure 5:   Closed chain of    three coupled anharmonic oscillators with quadratic potential ($ X_1^2, X_2^2,X_3^2$),  quartic potential ($ X_1^4, X_2^4,X_3^4$) and coupled interaction $  (X_1-X_2)^2,(X_2-X_3)^2,(X_3-X_1)^2$.  The two arrows form a closed chain.}
\\

In terms of the new coordinates\footnote{This is one of the orthogonal transformation of coordinates}  
\begin{align}
&X_1=  \frac{x_1}{\sqrt{3}}-\frac{x_2}{\sqrt{2}}+\frac{x_3}{\sqrt{6}},~~~ X_2=  \frac{x_1}{\sqrt{3}}+\frac{x_2}{\sqrt{2}}+\frac{x_3}{\sqrt{6}},~~~X_3=  \frac{x_1}{\sqrt{3}}-\sqrt{\frac{2}{3}} x_3 
\end{align}
the Hamiltonian is separated into kinetic energy, quadratic and quartic potential terms.
 \begin{align}
    H&= H^{(0)}+\lambda \ V_{\lambda}  \\
 H^{(0)}&={1\over 2}\Big[\( p_1^2+ p_2^2+p_3^2 \)+\( \tilde\omega^2x_1^2+\left( \tilde\omega ^2+3 \Omega ^2\right) x_2^2+\left( \tilde\omega ^2+3 \Omega ^2\right) x_3^2  \) \] \\
V _{\lambda} &= \left(\frac{2 x_1^4}{3}+x_2^4+x_3^4\right)+  \left(4  x_1^2x_2^2+ 2 x_2^2 x_3^2+4 x_3^2x_1^2\right) +\frac{4}{3} \sqrt{2}    \left(3x_1x_2^2 x_3-x_1 x_3^3 \right)  \label{5.5}
\end{align}
Hamiltonian $H^{(0)}$ describes the unperturbed harmonic energy and has a well known solution 
\be
&&H^{(0)} |n_1^{(0)},n_2^{(0)},n_3^{(0)} \>=E^{(0)}_n|n_1^{(0)},n_2^{(0)},n_3^{(0)} \>=\(  3\tilde\omega ^2+6 \Omega ^2 +{3\over2 }\) |n_1^{(0)},n_2^{(0),,n_3^{(0)}} \>  
\ee
which relates to eq.(\ref{3.2}) in the two site  system.  

 Now, there are {\it three unperturbed states} and, at first sight, by   the algorithms presented in the previous sections, we have to do  tedious calculations to find the OTOC property of the three coupled oscillators. In fact, using the property found in  2 coupled oscillators, we could argue that the  closed chain of  three coupled  oscillators will transition to quantum chaos phase at high temperature, like as that in the  2-coupled  oscillator system, and with Lyapunov exponent $\lambda_T\sim T^{1/4}$. The arguments are detailed below.
\\

First, we separate quartic potential term $V_\lambda$ into four parts
\be
V_{\lambda} =V_{12}+V_{23}+V_{31}+4 \sqrt{2} x_1x_2^2 x_3
\ee
where
\begin{align}
V_{12}&=   \frac{2 x_1^4}{3}+x_2^4  +   4  x_1^2x_2^2    ,~~~~~~~~~~~~~~~~~~~~V_{23}=   x_2^4+x_3^4 +    2 x_2^2 x_3^2 \nn\\
V_{31}&= x_3^4 +\frac{2 x_1^4}{3} + 4 x_3^2x_1^2 -\frac{4}{3} \sqrt{2} x_3^3  x_1 ,~~~~~V_{123}=4  \sqrt{2}   x_1x_2^2 x_3  \label{5.8}
\end{align}
Note that $V_{ij}$ are functions of coordinates $x_i,x_j$  while $V_{123}$ is function of coordinates $x_1,x_2,x_3$.  Then 
\begin{align}
H=H^{(0)}+\lambda V_{12}+\lambda V_{23}+\lambda V_{31} +\lambda V_{123}
\end{align}

Next, from the perturbation formulas, for example eq.(\ref{3.7}), the central quantity to be explicitly evaluated  becomes
\begin{align}
\<k^{(0)} |V|n^{(0)} \>&=\<k^{(0)} |V_{12}|n^{(0)} \>+\<k^{(0)} |V_{23}|n^{(0)} \>+\<k^{(0)} |V_{31}|n^{(0)} \>+\<k^{(0)} |V_{123}|n^{(0)} \>
\end{align}
We know that $|n^{(0)} \>=|n^{(0)}_1, n^{(0)}_2,n^{(0)}_3\>$ in which the quantum number $n^{(0)}_i$ is used to specified the harmonic oscillator at position "i", then  for example,  consider the first term in above equation   we have a simple result
\begin{align}
\<k^{(0)} |V_{12}|n^{(0)} \>&=\<k^{(0)}_1, k^{(0)}_2,k^{(0)}_3 | \left(\frac{2 x_1^4}{3}+x_2^4  +   4  x_1^2x_2^2\right)|n^{(0)}_1, n^{(0)}_2,n^{(0)}_3\>\\
&=\<k^{(0)}_1, k^{(0)}_2 | \left(\frac{2 x_1^4}{3}+x_2^4  +   4  x_1^2x_2^2\right)|n^{(0)}_1, n^{(0)}_2 \>\cdot \delta_{k^{(0)}_3, n^{(0)}_3}
\end{align}
In this way, the problem of  {\it three unperturbed states} reduces to the problem of {\it two unperturbed states} problem  in the two coupled anharmonic oscillators.

As  the potential of two coupled oscillator system $V= x_1^4+ x_2^4+6 x_1^2 x_2^2  $ in eq.(\ref{V1}) is now  replaced by $V_{12}$ the same algorithms could  be applied to study the OTOC property with interaction $V_{12}$, and terms of $V_{23},V_{31}$ too. The results show that each contribution leads to  the  same quantum chaos property, such as $\lambda_T\sim T^{1/4}$ with different critical temperatures, which is consistent  with the universality hypothesis of critical properties.\footnote{The property states that the critical properties of a system, specifically its critical exponents, depend fundamentally on the system global symmetry and its spatial dimension. Note that the precise value of the transition temperature (also called the critical temperature) $T_C$ which depends on the details of the system such as the coupling strength, is not  universal. } 

 Turn to the case with the potential  $V_{123}\sim  x_1x_2^2 x_3$ in eq.(\ref{5.8}).  The central quantity to be explicitly evaluated  can be expressed as
\begin{align}
\<k^{(0)} |V_{123}|n^{(0)} \>&\sim\<k^{(0)}_1, k^{(0)}_2,k^{(0)}_3 |  \left( x_1x_2^2 x_3\right)|n^{(0)}_1, n^{(0)}_2,n^{(0)}_3\>\\
&=\<k^{(0)}_1, k^{(0)}_2 | \left(x_1x_2^2\right)|n^{(0)}_1, n^{(0)}_2 \>\cdot   \< k^{(0)}_3 |  \left(   x_3\right)| n^{(0)}_3\>    \label{V123}
\end{align}
In this way, the problem of  {\it three unperturbed states} reduces to the problem of {\it two unperturbed states} problem  in the two coupled anharmonic oscillators and a single site system. 

Collecting these terms together we will see that the quantum chaos property in  a closed chain of  three coupled anharmonic oscillators has the same critical property as that in the  system  of  two coupled anharmonic oscillators.

\subsection{Closed chain of four coupled anharmonic oscillators}
We next investigate the closed chain of four coupled anharmonic oscillators. As before, from eq.(\ref{HH}) we have the Hamiltonian (in units where $M=1$) reads
\begin{align}
H&={1\over 2}\Big[\( P_1^2+P_2^2+P_3^2 +P_4^2\)+\tilde \omega^2(X_1^2+ X_2^2+X_3^2 )  \nn\\
&+ \Omega ^2\(   ( X_1- X_2)^2+ ( X_2- X_3)^2+ ( X_3- X_4)^2+ ( X_4- X_1)^2 \) 
& +2\lambda( X_1^4+X2^4+X_3^4 +X_4^4)\Big]    
\end{align} 
In terms of the new coordinates
\begin{align}
X_1&= \frac{x_1}{2}+\frac{x_2}{2}-\frac{x_3}{\sqrt{2}},~~~X_2= \frac{x_1}{2}-\frac{x_2}{2}-\frac{x_4}{\sqrt{2}},\nn\\
X_3&= \frac{x_1}{2}+\frac{x_2}{2}+\frac{x_3}{\sqrt{2}},~~~X_4= \frac{x_1}{2}+\frac{x_4}{\sqrt{2}}-\frac{x_2}{2}
\end{align}
the Hamiltonian is separated into kinetic energy, quadratic and quartic potential terms.
\begin{align}
    H&=H^{(0)}+\lambda \( V_{(I)} +\lambda \ V_{(IJ)} + \lambda \ V_{(IJK)}    \) \\
 H^{(0)}&={1\over 2}\[\( p_1^2+ p_2^2+p_3^2+p_4^2 \)++x_1^2 \omega ^2+x_2^2 \left(\omega ^2+4 \Omega ^2\right)+x_3^2 \left(\omega ^2+2 \Omega ^2\right)+x_4^2 \left(\omega ^2+2 \Omega ^2\right)\]\\
V_{(I)} &=\frac{3   x_1^4}{8}+\frac{3   x_2^4}{8}+   x_3^4+\frac{   x_4^4}{2}\\
  V_{(IJ)} &=
\frac{1}{2}    x_2 x_1^3-\frac{   x_4 x_1^3}{\sqrt{2}}+\frac{9}{4}    x_2^2 x_1^2+3    x_3^2 x_1^2+\frac{3}{2}   x_4^2 x_1^2+\frac{1}{2}    x_2^3 x_1-\sqrt{2}   x_4^3 x_1+\sqrt{2}   x_2 x_4^3+3    x_2^2 x_3^2\nn\\
&~~~+\frac{3}{2}   x_2^2 x_4^2+\frac{   x_2^3 x_4}{\sqrt{2}}\\
 V_{(IJK)} &=
\frac{3    x_2 x_4 x_1^2}{\sqrt{2}}+6    x_2 x_3^2 x_1-3    x_2 x_4^2 x_1-\frac{3    x_2^2 x_4 x_1}{\sqrt{2}}
\end{align}
where each term in $V_{(I )} $ depends on one   position, $V_{(IJ )} $ depends on two positions, $V_{(IJL )} $ depends on three positions.

To proceed, we  first separate term $V_{(I )} + V_{(IJ )} $ into below form
\begin{align}
V_{(I )} +V_{(IJ )} =V_{12}+V_{13}+V_{14}+V_{23}+V_{24}+V_{34}
\end{align}
where  $V_{ij}$ are function of coordinates  $x_i,x_j$.  Consider, for example, the potential term
\begin{align}
V_{12}=\frac{3   x_1^4}{8}+\frac{3   x_2^4}{8}+\frac{1}{2}  ( x_1^3 x_2 +x_1 x_2^3) 
\end{align}
We see that the potential of two coupled oscillator system $V= x_1^4+ x_2^4+6 x_1^2 x_2^2  $ in eq.(\ref{V1}) is now  replaced by above $V_{12}$, and therefore  the same algorithms could  be applied to study the OTOC property with interaction $V_{12}$, and terms of $V_{13},V_{14},V_{23},V_{24},V_{34}$ too. 

Finally, follow the arguments and schemes described in eq.(\ref{V123}) which investigate $  \<k^{(0)} |V_{123}|n^{(0)} \>$, we can evaluate  the parts of OTOC   from each term in  potenital  $V_{(IJK)}$.  Collect these results we will find that  a closed chain of four coupled anharmonic oscillators  exhibits  the quantum chaos as that in a coupled anharmonic oscillators.\footnote{Explicit calculations to confirm the conjecture are left for future work.}

In this way, the closed chain of N coupled anharmonic oscillators, which relates to the 1+1 dimensional $\lambda \phi^4$ theory, will have the same critical property as that in the  system  of  two coupled anharmonic oscillators.
 \section{Conclusions}
This paper investigates the out-of-time-order correlator (OTOC) in interacting quantum scalar field theory. We first regularize $\lambda \phi^4$ theory by discretizing it on a square lattice, which yields a system of coupled anharmonic oscillators.  We then  use the quantum mechanical method, which  was set up  by Hashimoto recently in \cite{Hashimoto17,Hashimoto20a,Hashimoto20b}, to study the OTOC of the coupled oscillator. 
Unlike prior studies that employed a wavefunction approach, we compute the OTOC using the second quantization method within a perturbative approximation.

 We first investigate the two coupled system and obtain  several  analytic relations of the spectrum, Fock space states, and matrix elements of the coordinate  to the second-order perturbation. We then use these relations to numerically analyze the associated thermal OTOC $C_T(t)$.  We find  that the function $C_T(t)$ exhibits the  exponential growth  fitting over a long time window  in the early stages with Lyapunov exponent $\lambda\sim T^{1/4}$,  which diagnose  quantum chaos. 

 Using these properties we furthermore investigate the closed chain of  3 and 4 coupled anharmonic  oscillators. We see that they have the same chaos property as that in 2 coupled anharmonic oscillators.  We argue that the property also shows in   N coupled anharmonic oscillators, which relates to the 1+1 dimensional interacting quantum scalar field system. 

In conclusion, an interesting property revealed in this paper is that signatures of quantum chaos emerge  at low orders of perturbative OTOC,  and a simple system of two coupled anharmonic oscillators already exhibits this behavior.   Finally, as the closed chain is a 1+1 dimensional system, it is useful to study the 1+2 system to examine how the properties of quantum chaos depend on spatial dimension. It would also be interesting to apply the prescription developed in this paper to investigate quantum chaos in other models, including those with fermions.

\newpage
\begin{center} 
{\bf  \large References}
\end{center}

\end{document}